\documentclass{article}
\usepackage{ijcai22}
\usepackage{xcolor}
\usepackage{times}
\usepackage{soul}
\usepackage{url}
\usepackage[hidelinks]{hyperref}
\usepackage[utf8]{inputenc}
\usepackage[small]{caption}
\usepackage{graphicx}
\usepackage{amsmath}
\usepackage{amsthm}
\usepackage{booktabs}
\usepackage{algorithm}
\usepackage{algorithmic}
\title{Systematic analysis of the effectiveness of adding human mobility data to covid-19 case prediction linear models}

\author{
Saad Mohammad Abrar$^1$\and
Naman Awasthi$^1$\and
Daniel Smolyak$^1$\And
Vanessa Frias-Martinez$^1$
\affiliations
$^1$University of Maryland, College Park
\emails
\{sabrar,nawasthi,dsmolyak,vfrias\}@umd.edu
}

\begin{document}

\maketitle

\begin{abstract}
Human mobility data has been extensively used in covid-19 case prediction models. 
Nevertheless, related work has questioned whether mobility data really helps that much.
We present a systematic analysis across mobility datasets and prediction lookaheads
and reveal that adding mobility data to predictive models improves model performance only for about two months at the onset
of the testing period, and that performance improvements - measured as predicted vs. actual correlation improvement over non-mobility baselines - are at most $0.3$. 
\end{abstract}
\vspace{-0.1cm}
\section{Introduction}
Human mobility data has been used in the past to model and characterize human behaviors in the built environment 
~\cite{vieira2010querying,hernandez2017estimating,frias2013cell,rubio2010human,wuspatial}, to support decision making for socio-economic development
~\cite{frias2010socio,fu2018identifying,frias2012mobilizing,hong2016topic,frias2012computing}, for public safety~\cite{wu2022enhancing,wu2023auditing}, as well as during epidemics and 
disasters~\cite{wesolowski2012quantifying,bengtsson2015using,hong2017understanding,isaacman2018modeling,ghurye2016framework,hong2020modeling}. 

The covid-19 pandemic has mainstreamed human mobility data into the public domain, and beyond academic networks. During the early stages of the pandemic, the importance of limiting human mobility to contain the epidemic became evident \cite{oh2021mobility,badr2020association}. Cities, states and countries took different non-pharmaceutical interventions (NPIs) some of which focused on mobility {\it e.g.,} from national lockdowns, to work-from-home approaches - whenever possible \cite{hsiang2020effect}. To evaluate the effect of these interventions, public health experts, the CDC, city departments and journalists explored the use of mobility data which, at the time, was made open and freely available by several for-profit companies. 
Companies like Apple, Google, Safegraph or Descartes shared different types of aggregated human mobility datasets to characterize features such as the volume of visits to specific places (e.g., schools, workplaces or restaurants), the volume of trips between regions (e.g., trips between two counties), or the volume of trips by type of transportation (e.g., driving vs. public transit). 

Beyond understanding the impact of mobility reduction policies, the increased access to mobility data sources has also supported research on covid-19 prediction models, with the assumption that how people moved - and interacted - in the past could provide additional information about how people get infected in the future. 
These case prediction models focus on providing estimates for future number of cases in the short- and in the long-term via lookahead analysis performance {\it i.e.,} measuring performance for various temporal windows such as next day, next week, or a month ahead \cite{ilin2021public,wang2020using}.
For example, 
researchers have shown that Safegraph data can help predict covid-19 cases at the county level in the US 
~\cite{nikparvar2021spatio}.

Despite the many predictive {\it success} stories using mobility data, there has been a small number of researchers that 
have shown that strong correlations between covid-19 cases and mobility do not hold after the onset of the pandemic \cite{Badr_2020,Gatalo_2020}. Although these works focused on correlation analyses, if translated to a prediction setting, these could imply that mobility data might help predict covid-19 cases early on during the pandemic, but fail to help as much in later periods, when other types of behaviors such as masking or social distancing were practiced, but were hard to incorporate into the models given the lack of granular behavioral spatial and temporal data across counties in the US.
Despite the importance of these findings, the community has not talked enough about them, probably due to the fact that papers looking into the temporal, long-term relationship between covid-19 cases and mobility data (i) only looked into correlations, rather than predictive models and (ii) only used one mobility dataset in their analyses \cite{Badr_2020,Gatalo_2020}. 
In addition, even if mobility data has proved to be helpful in correctly predicting covid-19 cases, related work often times has failed to compare mobility-based predictive results with non-mobility baseline models \cite{kuo2021evaluating,da2021meteorological} so as to quantify the improvements brought about by the mobility data itself. This is important given that mobility data comes at a cost to decision makers (Apple, Google, Safegraph or Descartes do not offer the data for free anymore).  

In this paper, we aim to carry out an in-depth, systematic analysis across types of mobility datasets and prediction lookaheads to (1) identify whether mobility data improves the performance of covid-19 case prediction models when compared to models that do not use mobility as a source of information; and (2) identify the temporal range when it does - if it does in fact help improve the prediction performance.

\begin{table*}[!htp]\centering
\scriptsize

\begin{tabular}{lll}\toprule
Dataset & County Counts & Description \\\midrule
New York Times Covid-19 
& 3135 & Daily case counts compiled from state and local governments.\\
Apple Mobility
& 2064 & Derived from Apple Maps, the mobility data is divided by transportation type $-$ we use their \textit{drive} category. \\
Descartes Mobility
& 2551 & Calculated median of maximum distances travelled by individuals based on geolocation data from mobile devices. \\
Google Mobility 
& 990 & Derived from Google Maps, the mobility data is divided by POI type $-$ we use their \textit{workplace} category. \\
Safegraph In/Intra/Out-flows
& 3116 & Daily volume of trips into (inflows), within (intraflows) and out of (outflows) each county, derived from mobile device geolocation data. \\
Safegraph POI
& 3065-3091 & Daily volume of trips to the following POIs: Grocery Stores, Religious Organizations, Restaurants, and Schools \\
\bottomrule
\end{tabular}
\caption{Descriptions of the covid-19 case dataset and each human mobility dataset, including number of counties available per dataset. }
\label{tab:data_description}

\end{table*}

\section{Methods}
In this section, we present the different components of our analysis  
to identify if mobility data improves the performance of covid-19 case prediction models when compared to  non-mobility data models; and, if it does, the temporal range for which mobility data is helpful {\it i.e.,} is it always helpful, or as prior work has pointed out, is it only helping at the onset? \cite{Gatalo_2020,Badr_2020}.
We also describe the covid-19 case and human mobility datasets as well as the predictive model we will use.

\subsection{Datasets}
We use ten human mobility datasets from four companies: safegraph \cite{safegraph}, google \cite{google}, apple \cite{apple}, and descartes \cite{warren2020mobility}; and covid-19 case data from the New York Times
\cite{nytimes_covid} (see Table \ref{tab:data_description} for details). Only counties with data available throughout the data collection period (03/2020-11/2020, 256 days) were considered in the analysis. To standardize the mobility features, we baseline each of the mobility datasets with respect to the average mobility level from February 17, 2020 to March 7, 2020 (to match the baseline used in the Descartes dataset), and then we take a rolling 7-day average (computed as the average of each mobility value and the values from the prior 6 days). We use a 7-day rolling average of the covid-19 case data as well.


\subsection{Predictive Models}
We focus on linear regression models due to their interpretability and common use among decision makers in the covid-19 case prediction context \cite{carvalho2019machine}. 
To evaluate the effectiveness of mobility data in the prediction of covid-19 cases over time, we analyze the performance of the linear regression using (1) only past covid-19 case data as independent variables to predict future covid-19 cases (non-mobility baseline), and (2) both covid-19 past case data and prior mobility data as independent variables; with the assumption that how people moved in the past could potentially provide additional information about how people get infected in the future. Both non-mobility baselines and mobility-based models are evaluated across five temporal prediction windows (a.k.a. lookaheads): 1-day, 7-days, 14-days, 21-days and 28-days. We estimate our models with elastic net regularization, as elastic net outperformed other regularization methods (OLS, lasso, and ridge) on average across all lookaheads.

We train one linear regression model per county and lookahead. The number of covid-19 cases for a given lookahead is predicted using the covid-19 cases for the prior lookahead value {\it e.g.,} for lookahead 1, the covid-19 cases for the prior day; and the 10-day lagged mobility, since the infection gap - from exposure to contagiousness - has been associated to that lag \cite{garcia2021improving}.
Lagged mobility data is only included in the mobility-based models that will be compared against the non-mobility baselines.

\subsection{Train and Test Methodology}

\begin{figure}
\centering
\includegraphics[scale=0.5]{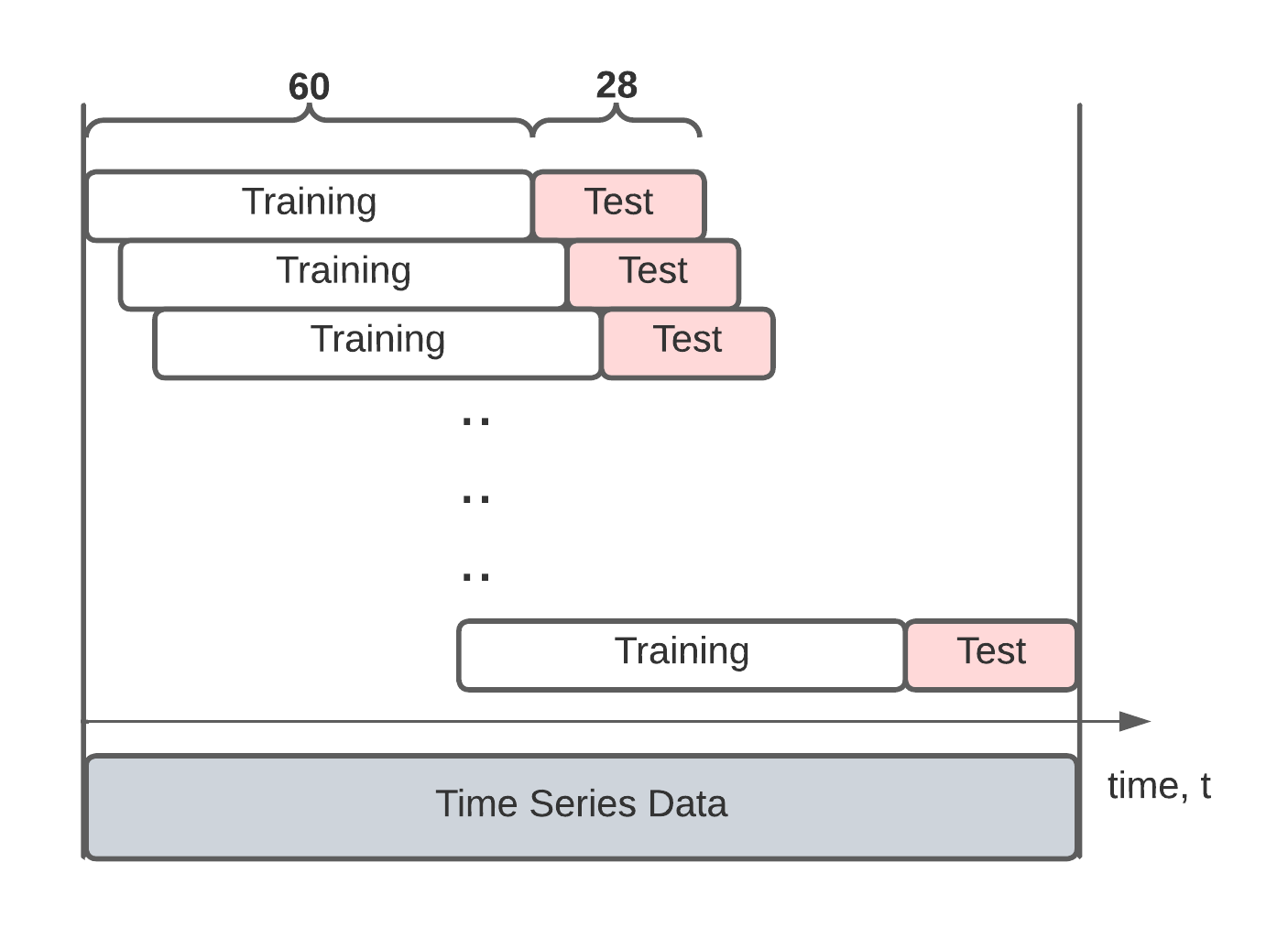}
\caption{Sliding-window approach for train and test data split.}
\label{fig:sliding}
\end{figure}

In order to identify whether mobility data improves the performance of covid-19 case prediction models when compared to their non-mobility counterpart, and if it does, identify the temporal range of improvement, we focus on small training and testing datasets. Specifically, we use the $8$ months of mobility data to create $60$-day training datasets and $28$-day testing datasets to test prediction accuracy for the next 28 days (lookaheads 1, 7, 14, 21 and 28). We construct each train-test pair using a sliding window as shown in Figure \ref{fig:sliding}). A total of $170$ train-test pairs where created with the available data. 

\subsection{Model Performance Analysis}


To identify 
whether mobility data improves the performance of covid-19 case prediction models when compared to their non-mobility counterpart, and if it does, identify the temporal range of improvement,
we propose the following approach. For each day in the testing period (170 days), we compute the difference in performance between mobility-based covid-19 case prediction models and their corresponding non-mobility baseline models for each lookahead and each human mobility dataset. Model performance is defined as the Spearman correlation between the number of predicted cases for each county in the dataset and their corresponding actual number of cases (ground truth covid-19 data). 

Formally, given a day $d$, lookahead $l$ and mobility dataset $m$, we compute the correlation improvement $ci$ as 
$$ci^m_{d,l} = \rho^{m}_{d,l} -\rho^{baseline}_{d,l}$$ 
with 
$ \rho^{m}_{d,l} = \rho (\{(p^{m}_{d,l,i},a_{d,l,i}) \mid i \in C \}) $
and
$ \rho^{baseline}_{d,l} = \rho (\{(p^{baseline}_{d,l,i},a_{d,l,i}) \mid i \in C \}) $
where $\rho$ is the Spearman correlation coefficient function, $C$ is the set of all counties for mobility dataset $m$, and $a$, $p^{m}$, and $p^{baseline}$ represent the actual covid-19 cases and the predicted cases for models that use mobility data or not, respectively. 

By looking at the daily correlation improvements over the testing period, we will be able to identify if mobility data improves the performance over non-mobility models, and if it does, evaluate the period of time when that happens. 
Positive values for the correlation improvement $ci$ indicate that mobility data improves the performance of covid-19 case prediction models when compared to models that do not use mobility data; while negative $ci$ indicate that baseline models that do not use mobility data perform better {\it i.e.,} mobility data hurts model performance.





\vspace{-0.25cm}
\section{Results}
\begin{figure*}
  \includegraphics[width=\textwidth,trim=3 3 3 3,clip]{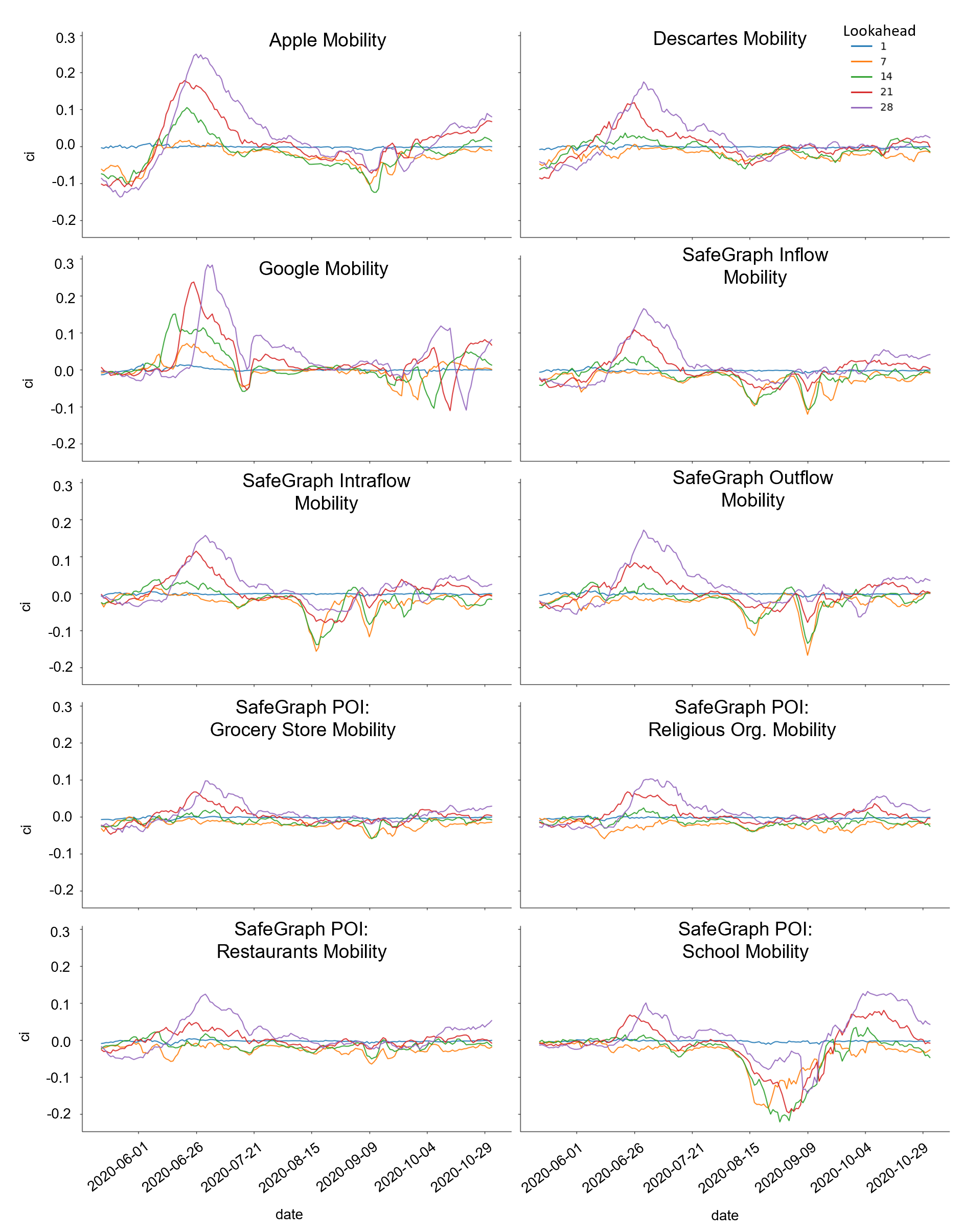}
  \caption{Spearman correlation improvement $ci$ of mobility models (apple, google, descartes, and safegraph inflows, intraflows, outflows, and safegraph POI (grocery stores, religious sites, restaurants, schools ) compared to baseline models. Each subplot shows five improvement trends, one for each lookahead $l$ on date $d$. }
    \label{fig:alternative1}
\end{figure*}

\vspace{-0.1cm}
We show all the changes in $ci$ over time across lookaheads and mobility datasets in Figure \ref{fig:alternative1}. 
Since $ci$ was defined as a subtraction, we cannot observe individual correlation values. However, it is important to clarify that the individual correlations for mobility and baseline models $\rho^{m}_{d,l}$ and $\rho^{baseline}_{d,l}$, respectively, were always positive {\it i.e., } an increase in the number of covid-19 cases was always associated to an increase in mobility, although at different levels of strength.  
To ease interpretation, we first discuss specific results for the apple mobility dataset and then proceed to discuss general trends across human mobility datasets. 

\begin{figure}
\centering
\includegraphics[width=0.95\linewidth,trim=3 3 3 3,clip]{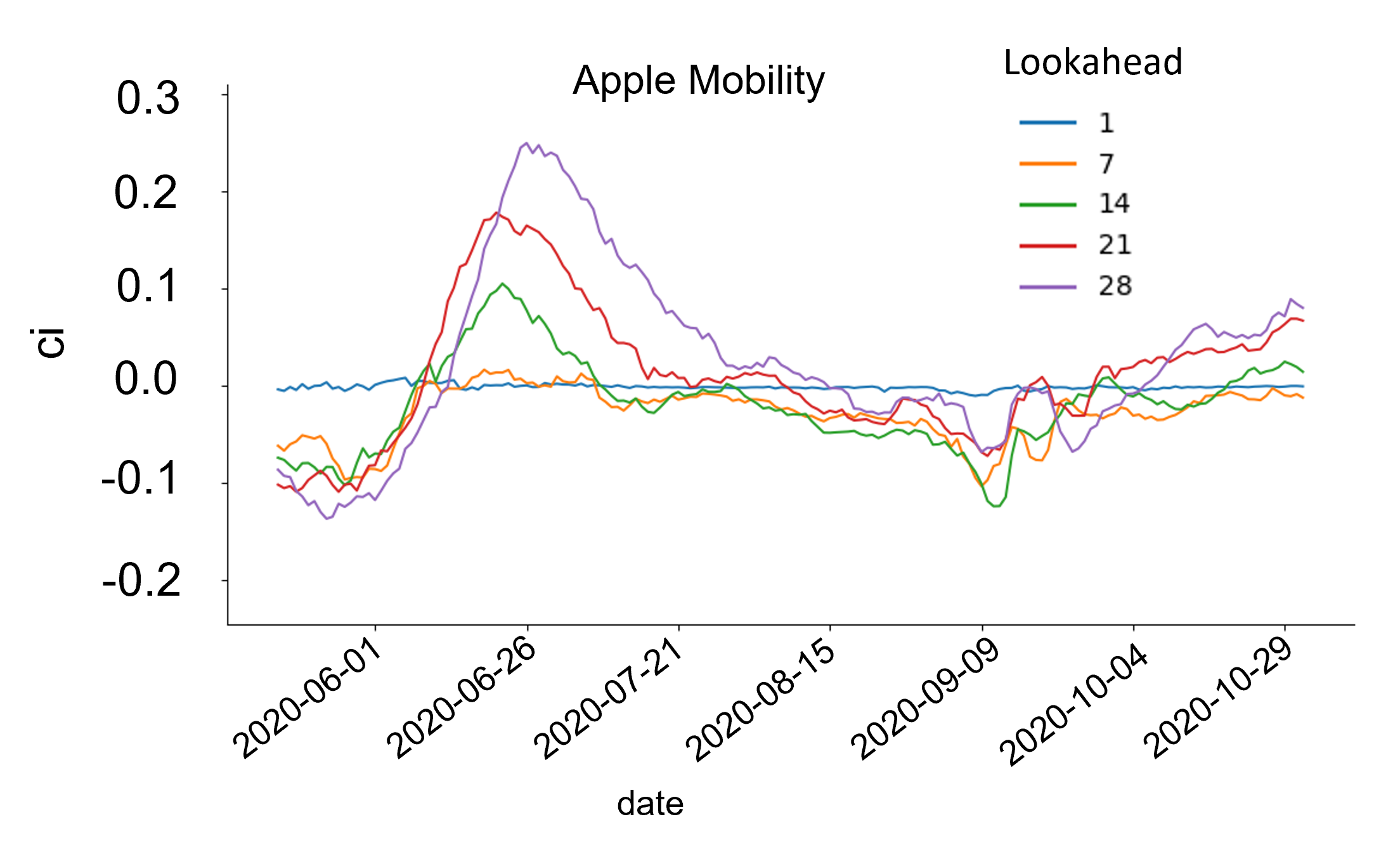}
\caption{Spearman correlation improvement $ci$ of apple mobility models compared to baseline models. Each subplot shows five improvement trends, one for each lookahead $l$ on day $d$.}
\label{fig:appleMob}
\end{figure}

\textbf{Apple Mobility.}
The correlation improvement plot for apple mobility is shown in Figure \ref{fig:appleMob}. The y-axis plots the correlation improvement $ci$ and the x-axis  the date $d$. 
Since our mobility data starts on March $18^{th}$ and we train for 60-day periods, our first testing day is May $17^{th}$.
Each colored line in the plot represents the correlation improvement over the baseline model (no mobility) for each lookahead $l$ considered in the analysis ($1,7,14,21,28$).
We observe the following:
\begin{enumerate}
\item There is no significant improvement for next day predictions (lookahead $l=1$) when mobility is added to the predictive model (orange line is close to 0).
\item For lookaheads $l>1$, and when adding apple mobility data helps ($ci>0$), we observe that the higher the lookahead, the larger the correlation improvement.  
\item For the first month, we see that baseline models perform better than models trained with apple mobility across all lookaheads (all $ci$ improvements are negative).
\item Addition of mobility to the model improves forecasts after $d \sim $ 8th June for around two months, with $ci$ values of up to $0.3$.
\item Baseline models perform comparably, or better, than mobility-based models using apple mobility between August-October.
\item After $d \geq$ 1st October, correlation improvements from addition of apple mobility data are positive again, but with very small values $ci \leq 0.1$.
\end{enumerate}

\textbf{Across Mobility Datasets.}
When analyzing trends across all mobility datasets, including apple (see Figure \ref{fig:alternative1}), we make the following observations: 

\begin{enumerate}
    \item All trends of correlation improvement described for apple mobility hold across the other datasets: for lookaheads $l>1$, correlation improvements are positive from mid-June to mid August ($ci \leq 0.3$), no improvements in August-October ($ci<0$), followed by minimal correlation improvements after October ($ci \leq 0.1$, with the exception of school mobility where improvements are large).   
    \item The maximum correlation improvement values are achieved with apple and google mobility data ($\approx 0.3$), followed by descartes and safegraph flows ($\approx 0.2$) and safegraph POI ($\approx 0.1$).
    
\end{enumerate}

\vspace{-0.25cm}
\section{Discussion and Future Work}

We have carried out a systematic analysis of the performance of mobility-based covid-19 case prediction models compared to their non-mobility baselines, across a diverse set of human mobility datasets and lookaheads. Our results show that 
adding mobility data to covid-19 case predictive models appears to significantly improve forecast performance over non-mobility baselines for about two months close to the onset of the testing period. After that, adding mobility either minimally improves or, even worse, hurts the performance of the prediction models. 
In addition, when mobility datasets improve covid-19 case prediction performance, it does so with maximum correlation improvements over the baseline of $ci \leq 0.3$, with larger values associated to higher lookahead predictions.   
In other words, augmenting covid-19 case prediction models with publicly available mobility datasets does not appear to extensively improve the predictive performance, or to improve it over long periods of time, raising questions about the effectiveness of mobility datasets and about effectiveness-cost trade offs, especially given the fact that the human mobility datasets used in this paper are not available for free anymore. 
Future work will expand this paper by looking into other predictive models including time-series models ARIMAX and Prophet and deep learning models like LSTM, GRU and Transformers. 
\vspace{-0.2cm}
\section*{Ethical Statement}
All mobility datasets and covid-19 case data have been retrieved from public data sources and are aggregated at the county level, minimizing privacy risks.

\appendix


\bibliographystyle{named}
\bibliography{ijcai22}

\end{document}